# Self-Partial and Dynamic Reconfiguration Implementation for AES using FPGA

Zine El Abidine ALAOUI ISMAILI and Ahmed MOUSSA

Innovative Technologies Laboratory,
National School of Applied Sciences,
Tangier, PBox 1818, Morocco
alaoui_zineabidine@yahoo.fr
amoussa@ensat.ac.ma

**Abstract**

This paper addresses efficient hardware/software implementation approaches for the AES (Advanced Encryption Standard) algorithm and describes the design and performance testing algorithm for embedded system.
Also, with the spread of reconfigurable hardware such as FPGAs (Field Programmable Gate Array) embedded cryptographic hardware became cost-effective. Nevertheless, it is worthy to note that nowadays, even hardwired cryptographic algorithms are not so safe.
From another side, the self-reconfiguring platform is reported that enables an FPGA to dynamically reconfigure itself under the control of an embedded microprocessor. Hardware acceleration significantly increases the performance of embedded systems built on programmable logic. Allowing a FPGA-based MicroBlaze processor to self-select the coprocessors uses can help reduce area requirements and increase a system's versatility.
The architecture proposed in this paper is an optimal hardware implementation algorithm and takes dynamic partially reconfigurable of FPGA. This implementation is good solution to preserve confidentiality and accessibility to the information in the numeric communication.

## 1. Introduction

Today, ultra deep submicronic technologies offer high scale density of integration for communication systems. This growth in integration has been accompanied with dramatically increase of complexity and transaction speed of this systems. As a consequence, security becomes a challenge and a critical issue especially for real time applications where materiel and software resources are very precious and necessary to provide a minimum of service quality.

Indeed, today speed and computing power impose the recourse to sophisticated and more complicated cryptography algorithms for high level security. Full software implementation is very heavy and slows down considerably speed of the information exchange. From another side, full hardware implementation is very expensive in terms of area, power and can also deteriorate speed of information transitions. This can be done dynamically at run-time and without user interaction, while the static part of the chip is not interrupted. The idea we put into practice is a coarse-grained partially dynamically reconfigurable implementation of a cryptosystem.

Our prototype implementation consists of a FPGA which is partially reconfigured at run-time to provide countermeasures against physical attacks. The static part is only configured upon system reset. Some advantages of dynamic reconfiguration for cryptosystems have been explored before [1, 2, 3]. In such systems, the main goal of dynamic reconfigurability is to use the available hardware resources in an optimal way. This is the first work that considers using a coarse-grained partially dynamically reconfigurable architecture in cryptosystems to prevent physical attacks by introducing temporal and/or spatial jitter [4, 5].

This paper presents an optimal implementation of the AES (Advanced Encryption Standard) cryptography algorithm by the use of a dynamic partially reconfigurable FPGA [6]. The reconfigurable aspect adapts the allowed basic bloc size to both the loop number and the size of the provided information, and makes all the AES blocs reconfigurable.
The paper is organized as follows: section 2 describes the AES algorithm. Reconfigurable FPGA and self reconfigurable methodology is presented in section 3, 4 and 5. The proposed methodology of algorithm implementation is given in section 6. Finally, results are presented and illustrated in section 7.

IJCSI

## 2. AES Encryption Algorithm

The National Institute of Standards and Technology (NIST) has initiated a process to develop a Federal Information Processing Standard (FIPS) for the AES, specifying an Advanced Encryption Algorithm to replace the Data Encryption Standard (DES) which expired in 1998 [6,7]. NIST has solicited candidate algorithms for inclusion in AES, resulting in fifteen official candidate algorithms of which five have been selected as finalists. Unlike DES, which was designed specifically for hardware implementations, one of the design criteria for AES candidate algorithms is that they can be efficiently implemented in both hardware and software. Thus, NIST has announced that both hardware and software performance measurements will be included in their efficiency testing. However, prior to the third AES conference in April 2000, virtually all performance comparisons have been restricted to software implementations on various platforms [5]. In October 2000, NIST chose Rijndael as the Advanced Encryption Algorithm.

The AES use the Rijndael encryption algorithm with cryptography keys of 128, 192, 256 bits. As in most of the symmetrical encryption algorithms, the AES algorithm manipulates the 128 bits of the input data, disposed in a 4 by 4 bytes matrix, with byte substitution, bit permutation and arithmetic operations in finite fields, more specifically, addition and multiplications in the Galois Field $2^8$ ($GF(2^8)$). Each set of operations is designated by round. The round computation is repeated 10, 12 or 14 times depending on the size of the key (128, 192, 256 bits respectively). The coding process includes the manipulation of a 128-bit data block through a series of logical and arithmetic operations. In the computation of both the encryption and decryption, a well defined order exists for the several operations that have to be performed over the data block.

The following describes in detail the operation performed by the AES encryption in each round. The State variable contains the 128-bit data block to be encrypted. In the Encryption part, first the data block to be encrypted is split into an array of bytes called as state matrix. This algorithm is based on round function, and different combinations of the algorithm are structured by repeating the round function different times. Each round function contains uniform and parallel four steps: SubBytes, ShiftRows, MixColumn and AddRoundKey transformation and each step has its own particular functionality. This is represented by this flow diagram. Here the round key is derived from the initial key and repeatedly applied to transform the block of plain text into cipher text blocks. The block and the key lengths can be independently specified to any multiple of 32 bits, with a minimum of 128 and a maximum of 256 bits. The repeated application of a round transformation state depends on the block length and the key length. For various block length and key length variable's value are given in table1.

The number of rounds of AES algorithm to be performed during the execution of the algorithm is dependent on the key size. The number of rounds, Key length and Block Size in the AES standard is summarized in Table 1 [8].

Table 1: Margin specifications Key-Block-Round Combinations for AES

|         | Key length (Nk round) | Key length (Bits) | Number of Round (Nr) |
|---------|-----------------------|-------------------|----------------------|
| AES-128 | 4                     | 128               | 10                   |
| AES-192 | 6                     | 192               | 12                   |
| AES-256 | 8                     | 256               | 14                   |

As mentioned before the coding process consists on the manipulation of the 128-bit data block through a series of logical and arithmetic operations, repeated a fixed number of times. This number of rounds is directly dependent on the size of the cipher key. In the computation of both the encryption and decryption, a well defined order exists for the several operations that have to be performed over the data block. The encryption/decryption process runs as follows in figure 1.

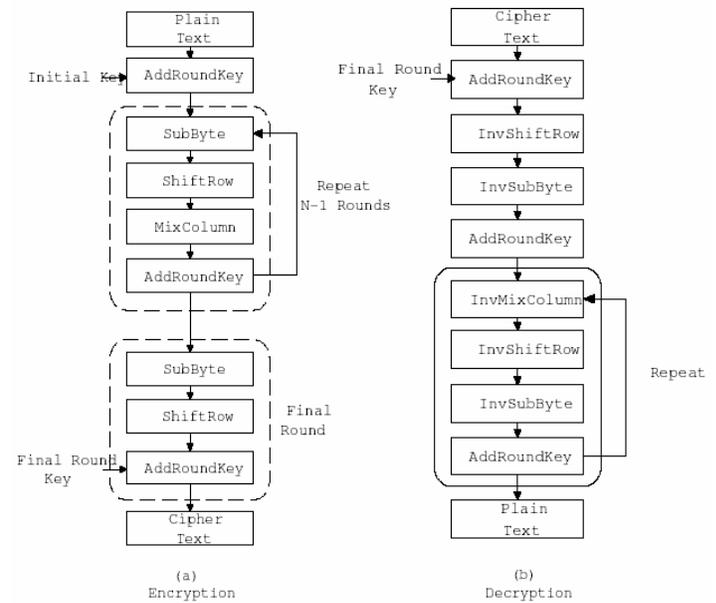

Fig.1: AES algorithm
(a) Encryption Structure   (b) Decryption Structure

The next subsections describe in detail the operation performed by each of the functions used above, for the particular case of the encryption.

## 2.1 The SubBytes Transformation

The SubBytes transformation is a non-linear byte substitution that acts on every byte of the state in isolation to produce a new byte value using an S-box substitution table. The action of this transformation is illustrated in Figure 2 for a block size of 4.
This substitution, which is invertible, is constructed by composing two transformations:
- First the multiplicative inverse in the finite field described earlier, with the {00} element mapped to itself.
- Second the affine transformation over $GF(2^8)$ defined by:

$$b'_i = b_i \oplus b_{(i+4) \bmod 8} \oplus b_{(i+5) \bmod 8} \oplus b_{(i+6) \bmod 8} \oplus b_{(i+7) \bmod 8} \oplus c_i \quad (1)$$

for $0 \leq i < 8$ where $b_i$ is the $i^{th}$ bit of the byte, and $c_i$ is the $i^{th}$ bit of a byte c with the value {63} or {01100011}. Here and elsewhere, a prime on a variable b' indicates that its value is to be updated with the value on the right.

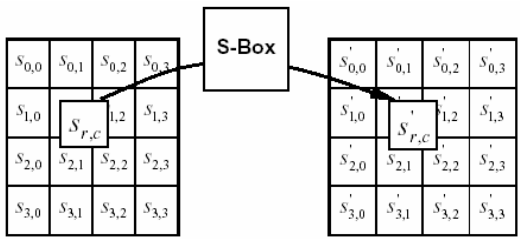

Fig. 2. SubBytes acts on every byte in the state

In matrix form the affine transformation element of this S-box can be expressed as:

$$\begin{bmatrix} b'_0 \\ b'_1 \\ b'_2 \\ b'_3 \\ b'_4 \\ b'_5 \\ b'_6 \\ b'_7 \end{bmatrix} = \begin{bmatrix} 1 & 0 & 0 & 0 & 1 & 1 & 1 & 1 \\ 1 & 1 & 0 & 0 & 0 & 1 & 1 & 1 \\ 1 & 1 & 1 & 0 & 0 & 0 & 1 & 1 \\ 1 & 1 & 1 & 1 & 0 & 0 & 0 & 1 \\ 1 & 1 & 1 & 1 & 1 & 0 & 0 & 0 \\ 0 & 1 & 1 & 1 & 1 & 1 & 0 & 0 \\ 0 & 0 & 1 & 1 & 1 & 1 & 1 & 0 \\ 0 & 0 & 0 & 1 & 1 & 1 & 1 & 1 \end{bmatrix} \begin{bmatrix} b_0 \\ b_1 \\ b_2 \\ b_3 \\ b_4 \\ b_5 \\ b_6 \\ b_7 \end{bmatrix} + \begin{bmatrix} 1 \\ 1 \\ 0 \\ 0 \\ 0 \\ 1 \\ 1 \\ 0 \end{bmatrix}$$

## 2.2 The ShiftRows Transformation

The ShiftRows transformation operates individually on each of the last three rows of the state by cyclically shifting the bytes in the row such that:

$$S'_{r,c} = S_{r,(c + shift(r, Nb)) \bmod Nb} \quad for\ 0 < r < 4\ and\ 0 \leq c \quad (2)$$

This has the effect of moving bytes to lower positions in the row except that the lowest bytes wrap around into the top of the row (note that a prime on a variable indicates an updated value). The action of this transformation is illustrated in Figure 3.

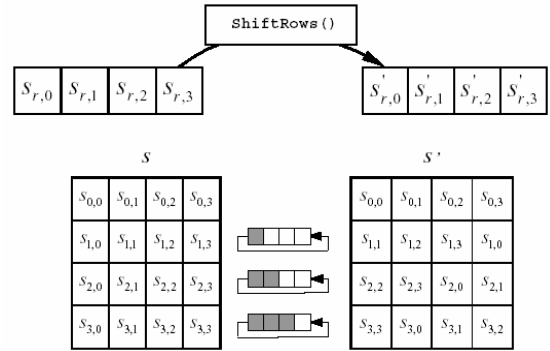

Fig. 3: Proposed beam former ShiftRows() cyclically shifts the last three rows in the State.

## 2.3 The MixColumns Transformation

The MixColumns transformation acts independently on every column of the state and treats each column as a four-term polynomial. The columns are considered as polynomials over $GF(2^8)$ and multiplied modulo $x^4+1$ with a fixed polynomial a(x), given by

$$a(x) = \{03\}x^3 + \{01\}x^2 + \{01\}x + \{02\} \quad (3)$$

This equation can be written as a matrix multiplication. Let:

$$S'(x) = a(x) \times S(x): \quad (4)$$

In matrix form the transformation used given in where all the values are finite field elements as discussed in Section 2.

$$\begin{bmatrix} S'_{0,c} \\ S'_{1,c} \\ S'_{2,c} \\ S'_{3,c} \end{bmatrix} = \begin{bmatrix} 02 & 03 & 01 & 01 \\ 01 & 02 & 03 & 01 \\ 01 & 01 & 02 & 03 \\ 03 & 01 & 01 & 02 \end{bmatrix} \begin{bmatrix} S_{0,c} \\ S_{1,c} \\ S_{2,c} \\ S_{3,c} \end{bmatrix} \quad for\ 0 \leq c \quad (5)$$

The action of this transformation is illustrated in Figure 3.

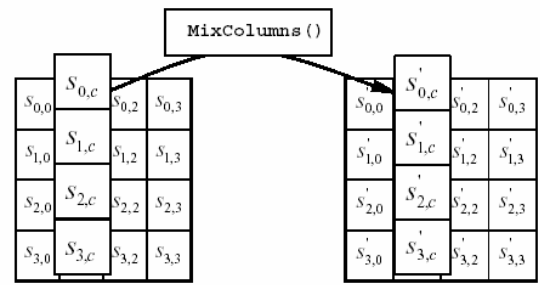

Fig. 4: MixColumns() operates on the State column-by-column.

## 2.4 The AddRoundKey Transformation

In the AddRoundKey transformation Nb words from the key schedule, described later, are each added (XOR) into the columns of the state so that:

$$[V'_{0,c}, V'_{1,c}, V'_{2,c}, V'_{3,c}] =$$
$$[b_{0,c}, b_{1,c}, b_{2,c}, b_{3,c}] \oplus [w_{round*Nb+c}] \; for \; 0 \leq c \leq Nb \quad (6)$$

Where the key schedule words [7] will be described later and round is the round number in the range $1 \leq round \leq Nr$. The round number starts at 1 because there is an initial key addition prior to the round function. The action of this transformation is illustrated in Figure 5.

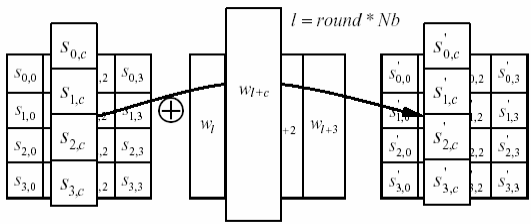

Fig.5: AddRoundKey() XORs each column of the State with a word from the key schedule.

## 3. Reconfigurable Hardware Technology

Field Programmable Gate Array (FPGA) is an integrated circuit that can be bought off the shelf and reconfigured by designers themselves. With each reconfiguration, which takes only a fraction of a second, an integrated circuit can perform a completely different function. FPGA consists of thousands of universal building blocks, known as Configurable Logic Blocks (CLBs), connected using programmable interconnects. Reconfiguration is able to change a function of each CLB and connections among them, leading to a functionally new digital circuit.

In recent years, FPGAs have been used for reconfigurable computing, when the main goal is to obtain high performance at a reasonable coast at the hardware implemented algorithms. The main advantage of FPGAs is their reconfigurability, i.e. they can be used for different purposes at different stages of computation and they can be.

Besides Cryptography, application of FPGAs can be found in the domains of evolvable and biologically-inspired hardware, network processor, real-time system, rapid ASIC prototyping, digital signal processing interactive multimedia, machine vision, computer graphics, robotics, embedded applications, and so forth. In general, FPGAs tend to be an excellent choice when dealing with algorithms that can benefit from the high parallelism offered by the FPGA fine grained architecture.

Significant technical advances have led to architecture to combine FPGAs logic blocks and interconnect matrices, with one or more microprocessors and memory blocks integrated on a single chip [9, 10]. This hybrid technology is called Configurable System on Chip (CSoC). Example for the CSoC technology are the Xilinx Virtex Pro II, the virtex 4, and virtex 5 FPGAs families, with include one or more hard-core Power PC processor embedded along with the FPGA's logic fabric.

Alternatively, soft processor cores that are implemented using part of the FPGAs logic fabric are also available. This approach is more flexible and less costly than the CSoC technology [11]. Many soft processors core are now available in commercial products. Some of the most notorious examples are: Xilinx 32-bits MicroBlaze and PicoBlaze, and the Altera Nios and 32-bits Nios II processors. These soft processor cores are configurable in the since that the designer can introduce new custom instructions or data paths. Furthermore, unlike the hard-core processors included in the Configurable System-on-Chip (CSoC) technology, designers can add as many soft processor cores as they may need. (Some designs could include 64 such processors or even more).

## 4. Dynamic Partial Reconfiguration

The incredible growth of FPGA capabilities in recent years and the new features included on them has opened many new investigation fields. One of the more interesting ones concerns partial reconfiguration and its possibilities [12,9]. This feature allows the device to be partially reconfigured while the rest of the device continues its normal operation. Partial reconfiguration is the ability to reconfigure preselected areas of an FPGA anytime after its initial configuration while the design is operational. By taking advantage of partial reconfiguration, hardware can be shared between various applications and upgraded remotely without rebooting and thus resource utilization can be increased [12].

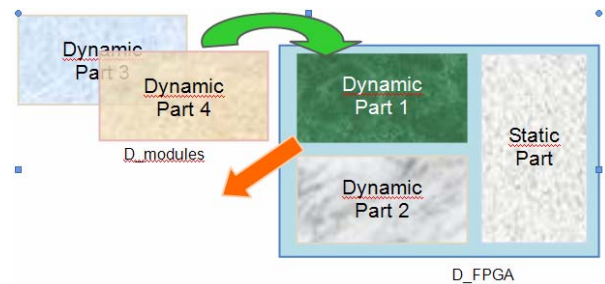

Fig. 6: Reconfigurable FPGA structure

FPGA devices are partially reconfigured by loading only a subset of configuration frames into the FPGA internal configuration memory. The Xilinx Virtex-II Pro FPGAs allow partial reconfiguration in two forms: static and dynamic.

Static (or shutdown) partial reconfiguration takes place when the rest of the device is inactive and in shutdown mode. The non-reconfigurable area of the FPGA is held in reset and the FPGA enters the start-up sequence after partial reconfiguration is completed. In contrast, in dynamic (or active) partial reconfiguration new data can be loaded to dynamically reconfigure a particular area of FPGA while the rest of it is still operational. User design is not suspended and no reset and start-up sequence is necessary.

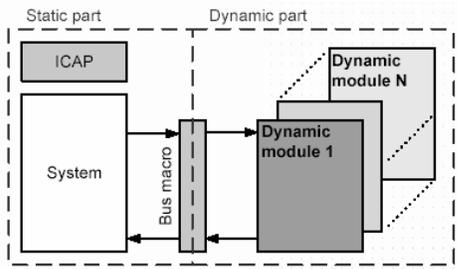

Fig. 7: Static and dynamic part for system reconfigurable

## 5. Self Partial Dynamic Reconfiguration

The Dynamic Partial Self-Reconfiguration (DPSR) concept is the ability to change the configuration of part of an FPGA device by itself while other processes continue in the rest of the device. A self-reconfiguring platform is reported that enables an FPGA to dynamically reconfigure itself under the control of an embedded microprocessor [10].

A partially reconfigurable design consists of a set of full designs and partial modules. The full and partial bitstreams are generated for different configurations of a design. The idea of implementing a self-reconfiguring platform for Xilinx Virtex family was first reported in [10]. The platform enabled an FPGA to dynamically reconfigure itself under the control of an embedded microprocessor.

The hardware component of Self Reconfiguring Platform (SRP) is composed of the internal configuration access port (ICAP), control logic, a small configuration cache, and an embedded processor. The embedded processor can be Xilinx Microblaze, which is a 32-bit RISC soft processor core [13]. The hard-core Power PC on the virtex II Pro can also be used as the embedded processor. The embedded processor provides intelligent control of device reconfiguration run-time. The provided hardware architecture established the framework for the implementation of the self-reconfiguring platforms. Internal configuration access port application program interface (ICAP API) and Xilinx partial reconfiguration toolkit (XPART) provide methods for reading and modifying select FPGA resources and support for re-locatable partial bitstreams.

Taking advantage of FPGA capacity presented above, we try to develop a flexible architecture of the AES implementation. The complexity of this arises from the algorithm architecture associated to the loop number and information size [13,4, 5, 14, 15, 3, 16].

The main idea of this work is to adapt the basic bloc size to the loop number and the size of the available information. The global architecture of the proposed system using dynamically reconfigurable FPGA is illustrated below (Cf. Fig. 8)

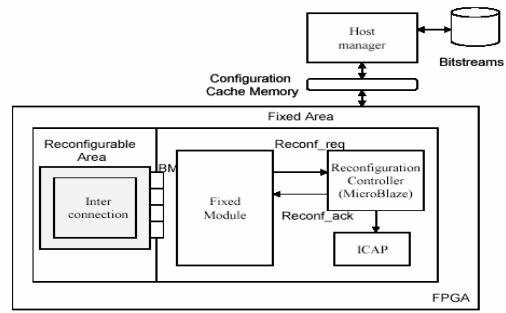

Fig. 8: Global architecture for self-reconfigurable system

## 6. A Self-Reconfigurable Implementation of AES

Our principal contribution in this article, is to conceive an optimal system allowing the implementation of the AES by using the self-reconfigurable dynamic method.

### 6.1 Methodology implementation

To increase the performance of the implemented circuit, especially cost, power and inaccessibility, all of the AES blocs may be reconfigurable [17]. So, the used parameters for reconfiguration are implanted inside the manager module of reconfiguring, and it is possible to quickly cross from a safe configuration to another by updating a hard system protection.

The control and management module of reconfiguration allows choosing a correct memory program (PR) and generating a reconfiguration signal (Cf. Fig. 9). The real dynamic reconfiguration procedure of the AES is preceded by two controllers: the first one, achieved by Microblaze processor, computes the reconfiguration parameters using the available signal and the key size. This is the current state of the system. The second one computes the best parameters under input constraints, and writes these parameters in the configuration register for managing the reconfiguration process.

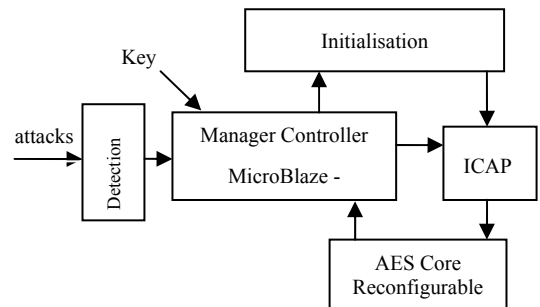

Fig.9: Global architecture for implementation the AES

In Figure 10, we present the modular design and reconfiguration cryptosystem.

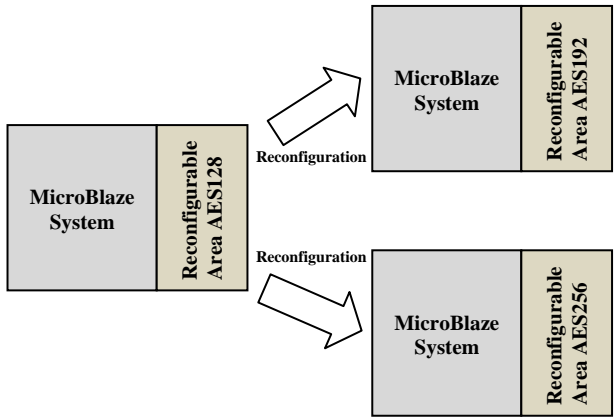

Fig. 10: Modular Design and Reconfiguration cryptosystem.

### 6.2 Configuration controller finite state machine

As described previously, the configuration controller is developed with a finite state machine. With the knowledge of the memory mapping, the configuration management finite state machine is relatively simple. The configuration controller is used only for normal FPGA configuration when power is switched on.

Figure 11 shows the four-global-states used by the configuration controller.

The first state of this four-states FSM (Finite State Machine) is an start state. To change state the configuration controller waits for detection the length key signal. This signal is the begin-signal of the normal configuration process.

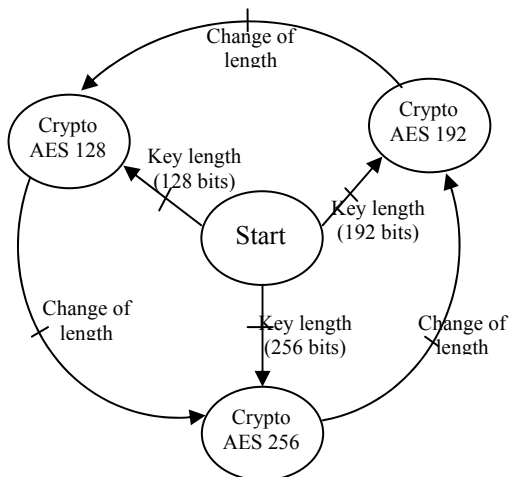

Fig. 11: Finite state machine configuration controller

## 7 Implementation Result

To test the proposed method, in first time, we have implemented the AES algorithm with on the Spartan II (XC2S200E) and Virtex II (XC2V500) of Xilinx. The results are summarized in the Table 2.

Table 2: comparison of the different implementations of the AES

| | FPGA Resource | Resource Used/Total Resource (XC2S200E) | Resource Used/Total Resource (XC2V500) |
|---|---|---|---|
| AES-128 | Slices | 196/2353 | 192/3072 |
| | Slice Flip-Flops | 92/4704 | 78/6144 |
| | 4-input LUTs | 352/4704 | 342/6144 |
| | BRAMs | 6/14 | 6/32 |
| AES-192 | Slices | 265/2353 | 241/3072 |
| | Slice Flip-Flops | 102/4704 | 76/6144 |
| | 4-input LUTs | 467/4707 | 341/6144 |
| | BRAMs | 6/14 | 6/32 |
| AES-256 | Slices | 252/2353 | 207/3072 |
| | Slice Flip-Flops | 99/4704 | 81/6144 |
| | 4-input LUTs | 469/4704 | 381/6144 |
| | BRAMs | 6/14 | 6/32 |

The performance implementation of AES cryptographic is presented in the table 3.

Table 3: Performance implementation for AES

| | Parameter | Device XC2S200E | Device (XC2V500) |
|---|---|---|---|
| AES-128 | Minimum Period (ns) | 35.520 | 13.674 |
| | Maximum Frequency | 28.742 | 78.59 |
| | Clock Cycle Used | 250 | 250 |
| | Thtoughput (Mbps) | 16.362 | 40.57 |
| | TPS (kbps/slice) | 83 | 232 |
| AES-192 | Minimum Period (ns) | 41.387 | 13.863 |
| | Maximum Frequency | 25.825 | 71.78 |
| | Clock Cycle Used | 300 | 300 |
| | Thtoughput (Mbps) | 11.361 | 31.72 |
| | TPS (kbps/slice) | 41 | 135 |
| AES | Minimum | 37.648 | 15.043 |

| | | |
|---|---|---|
| Period (ns) | | |
| Maximum Frequency | 27.067 | 70.975 |
| Clock Cycle Used | 350 | 350 |
| Throughput (Mbps) | 9.739 | 26.734 |

After checking of different hardware implementation of algorithms from the AES, we passed to the total test of the system of self reconfiguration a base the Microblaze processor, the results of this implementation in virtex II pro is shown on the table 4.

We notice that one can easily pass from a configuration to another using the software program implemented in the processor Microblaze.

As described previously, the configuration controller is developed with a finite state machine in figure 11. With the knowledge of the memory mapping, the configuration management finite state machine is relatively simple.

Table 4: Implementation of Microblaze and cryptosystem

| | | FPGA Slices | LUTs | FF/Latches | BRAM |
|---|---|---|---|---|---|
| | MicroBlaze System | 4083 | 3383 | 3228 | 25 |
| AES coprocessor | AES 128 | 3565 | 3086 | 3042 | 4 |
| | AES-192 | 3764 | 3259 | 3149 | 4 |
| | AES-256 | 3632 | 3127 | 3205 | 4 |

## 8 Conclusion

In this paper we present the AES coprocessor implementation using the self partial dynamically reconfiguration of FPGA. The main advantage of this works appear in the capacity of the proposed architecture to modify or/and change the size of the key without stopping the normal operation of the system. As a consequence, the proposed system is able to increase the security and safety of the AES algorithm.

Moreover, implementation of the AES crypto-processor with this new configuration illustrates the ability of this architecture to optimize the processor occupation and the reconfiguration time.

In order to explore the encoding method on the self-partial dynamic reconfiguration, our short-term prospect, in the feature work, consists with the implementation of this algorithm in a real communication system.

## References

[1] F.-X. Standaert, G. Rouvroy, J.-J. Quisquater and J.-D. Legat, "Efficient implementation of Rijndael encryption in reconfigurable hardware: Improvements and design tradeoffs," **in the proceedings of CHES 2003, Lecture Notes in Computer Science**, Cologne Germany September 2003, pp. 334–350.

[2] Ming-Haw Jing, Zih-Heng Chen, Jian-Hong Chen, and Yan-Haw Chen, "Reconfigurable system for high speed and diversified AES using FPGA", **Microprocessors and Microsystems,** vol. 31, Issue 2, March 2007, pp. 94-102.

[3] A.J Elbirt., W. Yip, B. Chetwynd, C. Paar "An FPGA-based performance evaluation of the AES block cipher candidate algorithm ", **IEEE Transactions on Very Large Scale Integration (VLSI) Systems**, vol. 9 Issue 4, 2001, pp. 545 – 557.

[4] M. McLoone and J.V.McCanny: "High Performance Single-Chip FPGA Rijndael Algorithm Implementations", **Cryptographic Hardware and Embedded Systems** (CHES 2001), Paris, France, 2001.

[5] National Institute of Standards and Technology (NIST), **Second Advanced Encryption Standard (AES) Conference**, Rome, Italy, March 1999.

[6] B. Schneier, "Applied Cryptography", **John Wiley & Sons Inc.**, New York, USA, 2nd ed., 1996.

[7] M. Kandemir, W. Zhang, and M. Karakoy, "Runtime code parallelization for onchip multiprocessors", **In Proceedings of the 6th Design Automation and Test in Europe Conference**, Munich, Germany, March, 2003.

[8] J. Daemen, V. Rijmen : "AES Proposal: Rijndael, The Rijndael Block Cipher", **AES Proposal**, 1999, pp. 1–45.

[9] M. Huebner, C. Schuck, M. Kuhnle, and J. Becker, "New 2-Dimensional Partial Dynamic Reconfiguration Techniques for Real-time Adaptive Microelectronic Circuits," **Proc. Of Emerging VLSI Technologies and Architectures**, Karlsruhe, Germany ,Mars 2006.

[10] Xilinx web site. http://www.xilinx.com/ipcenter/processorcentral/microblaze (2003).

[11] P. Lysaght, B. Brodget, J. Mason, J. Young, and B. Bridgford, "Enhanced Architectures, Design Methodologies and CAD Tools for Dynamic Reconfiguration of Xilinx FPGAs", **International Conference on Field Programmable Logic and Applications**, Madrid, Spain, 2006.

[12] M. Ullmann, M. Huebner, B. Grimm, and J. Becker, "An FPGA Run-Time System for Dynamical On-Demand Reconfiguration," **Proc. of the 18th International Parallel and Distributed Processing Symposium,** Karlsruhe, Germany April 26-30, 2004.

[13] H. Qin, T. Sasao and Y. "An FPGA Design of AES Encryption Circuit with 128-bit Keys" **Proceedings of the 15th ACM Great Lakes symposium on VLSI**, Chicago, Illinois, USA, April 17–19, 2005,.

[14] O.Perez, Y.Berviller,C.Tanougast, and S.Weber, "The Use of Runtime Reconfiguration on FPGA Circuits to Increase

the Performance of the AES Algorithm Implementation", **Journal of Universal Computer Science**, vol. 13, no. 3, 2007, pp.349-362.

[15] N. Saqib, F.Rodriguez-Henriquez, and A. Diaz-Pérez, "Two approaches for a single-Chip FPGA Implementation of an Encyptor/Decryptor AES Core" **International C-Conference on Field-Programmable Logic and Applications**, Lisbon, Portugal, September 2003.

[16] M Mogollon: "Cryptography and Security Services: Mechanisms and Applications" **Cybertech Publishing**, 2007.

[17] Z.A, Alaoui, A. Moussa, A. Elmourabit and K. Amechnoue "Flexible Hardware Architecture for AES Cryptography Algorithm" **IEEE Conference on Multimedia Computing and Systems**, ouarzazate, morocco, April 2009.

**Z. alaoui-Ismaili**, received the DEA in electronics in 1997 and the Ph.D. degree in Electronics and industrial Computer Engineering in 2002, both from University IbnTofail de Kenitra, Morocco. He is currently researcher teacher at the Telecoms & Electronics department of National School of Applied Sciences tangier, Morocco, since June 2003.
His main research interests are FPGA based reconfigurable computing applications, with a special focus on dynamic partial reconfiguration and embedded systems.
Dr. Alaoui_Ismaili authored or coauthored more than 10 papers journal and conference.
He is president of Association Moroccan Society of microelectronics.

**A. Moussa,** was born in 1970 in Oujda, Morocco. He received the Licence in Electronics from the University of Oujda, Morocco, in 1994, and the PhD in Automatic Control and Information Theory from the University of Kenitra, Morocco, in 2001. He worked two years as a post-graduate researcher at the University of Sciences and Technology of Lille, France. At 2003 he joined Sanofi-Aventis research laboratory in Montpellier, France where he supervised Microarray analysis activities .He is now a professor at the National School of Applied Sciences in Tangier-Morocco and his current research interests are in the application of the Markov theory and multidimensional data analysis to image processing, and embedded systems.